\pdfoutput=1

\documentclass[11pt]{article}

\usepackage[table]{xcolor}
\usepackage[]{acl}

\usepackage{times}
\usepackage{latexsym}

\usepackage[T1]{fontenc}

\usepackage[utf8]{inputenc}

\usepackage{color}
\usepackage{tcolorbox}
\usepackage{tabularx}
\usepackage{tabu}
\usepackage{booktabs}
\usepackage{multirow}

\usepackage{graphicx} 
\usepackage{float}
\usepackage{subfigure} 
\usepackage{amssymb}

\usepackage{hyperref}
\usepackage{tablefootnote}
\usepackage{xspace}

\usepackage{float}
\usepackage{amsmath}
\usepackage{bm}
\usepackage{algorithmicx}
\usepackage{algorithm}
\usepackage{algpseudocode}

\usepackage{arydshln}

\usepackage{amssymb}
\usepackage{pifont}
\usepackage{enumitem}

\usepackage{microtype}

\newcommand{\ours}{TestCase-Eval\xspace}
\newcommand{\cf}{Codeforces\xspace}
\newcommand{\eg}{\hbox{\emph{e.g.,}}\xspace}
\newcommand{\ie}{\hbox{\emph{i.e.,}}\xspace}
\newcommand{\nproblem}{500\xspace}
\newcommand{\nsubmission}{100,000\xspace}
\newcommand{\nmodel}{19\xspace}

\newcommand{\huggingface}{\raisebox{-1.5pt}{\includegraphics[height=1.05em]{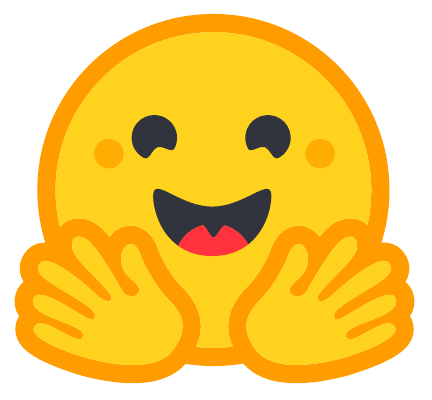}}\xspace}
\newcommand{\github}{\raisebox{-1.5pt}{\includegraphics[height=1.05em]{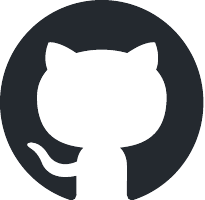}}\xspace}

\title{Can LLMs Generate High-Quality Test Cases for Algorithm Problems? \\\ours: A Systematic Evaluation of Fault Coverage and Exposure}

\author{
Zheyuan Yang$^\clubsuit$ \quad 
Zexi Kuang$^\diamondsuit$ \quad 
Xue Xia$^\heartsuit$ \quad 
Yilun Zhao$^\spadesuit$  
\vspace{5pt}
\\
$^\clubsuit$Tongji University \quad
$^\diamondsuit$Northeastern University \quad
$^\heartsuit$HKUST \quad 
$^\spadesuit$Yale University \quad 
}

\begin{document}
\maketitle
\begin{abstract}
We introduce \ours, a new benchmark for systematic evaluation of LLMs in test-case generation. 
\ours includes \nproblem algorithm problems and \nsubmission human-crafted solutions from the Codeforces platform.
It focuses on two pivotal tasks: 
(1) \emph{Fault Coverage}, which measures how well LLM-generated test sets probe diverse input scenarios and cover a wide range of potential failure modes. 
(2) \emph{Fault Exposure}, which evaluates whether LLMs can craft a tailored test input that reveals a specific incorrect code implementation.
We provide a comprehensive assessment of \nmodel state-of-the-art open-source and proprietary LLMs on \ours, offering insights into their strengths and limitations in generating effective test cases for algorithm problems.

\vspace{-15pt}
\begin{small}
\begin{center}
\begin{tabular}{cll}
\huggingface & \textbf{Data} & \href{https://huggingface.co/TestCase-Eval} {\path{TestCase-Eval}}\\
\github & \textbf{Code} & \href{https://github.com/FlowRays/TestCase-Eval}{\path{FlowRays/TestCase-Eval}}\\
\end{tabular}
\end{center} 
\end{small}
\vspace{5pt}

\end{abstract}

\section{Introduction}
Algorithmic problem-solving is fundamental to computational fields such as software engineering, data science, and competitive programming~\cite{Jimenez2023SWEbenchCL, Huang2023MLAgentBenchEL, Jain2024LiveCodeBenchHA, yu2024humanevalprombpppro, ElKishky2025CompetitivePW}. 
The correctness and robustness of algorithmic solutions hinge on the quality of test suites—carefully designed inputs that uncover edge cases, corner conditions, performance limitations, and common failure scenarios~\cite{Austin2021ProgramSW, Hendrycks2021MeasuringCC, Li2022CompetitionlevelCG}. 
Traditionally, crafting such test cases requires significant domain expertise and manual effort. With the rapid advancement of LLMs capable of sophisticated code generation, a crucial question arises: \textbf{Can LLMs generate high-quality test cases that match or surpass those designed by human experts?}

\begin{figure*}[!t]
\centering
\includegraphics[width=1\linewidth]{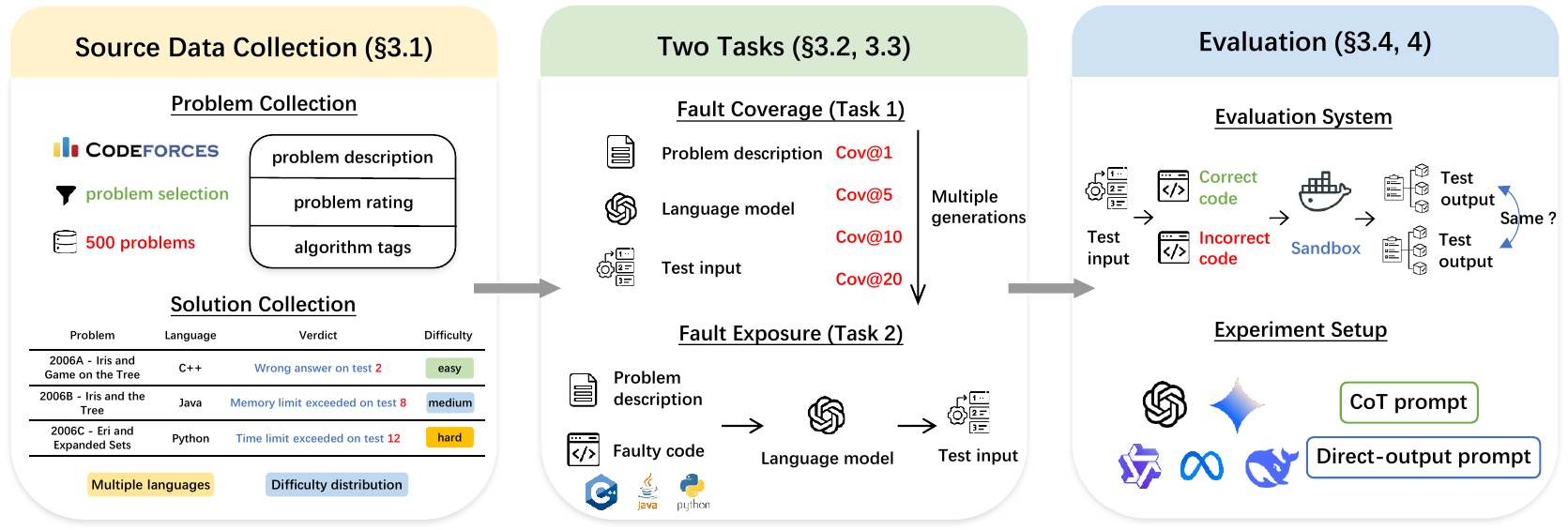}
\caption{
An overview of \ours and the research pipeline in this study.
}
\label{fig:main-example}
\end{figure*}

We introduce \ours,
a comprehensive benchmark for systematically evaluating LLMs in test-case generation for algorithmic problems. 
It comprises \nproblem up-to-date algorithm problems and \nsubmission corresponding real-human crafted solutions, both sourced from the Codeforces platform. 
As illustrated in \autoref{fig:main-example}, \ours features two core tasks, each targeting a crucial aspect of test-case quality: 
(1) \emph{Fault Coverage}, which evaluates whether LLM-generated test cases effectively explore diverse input scenarios, including edge cases and boundary conditions, to expose various types of incorrect solutions.
(2) \emph{Fault Exposure}, which evaluates whether an LLM can generate a targeted test input that successfully exposes the flaws in a specific given incorrect solution.

We conduct an extensive evaluation on \ours, covering \nmodel frontier 
open-source and proprietary LLMs. 
Our experimental results demonstrate that \ours presents a significant challenge, with even top-performing models like Qwen3-32B scoring only 43.8\% on the Fault Exposure task—far below human expert performance (93.3\%). These findings underscore the inherent difficulty of \ours. Furthermore, our in-depth analysis of reasoning LLMs, CoT reasoning, and model performance across different programming languages and error types offers valuable insights for future advancements in the field.

\section{Related Work}
Prior works on LLM-based test-case generation follows two main directions:
(1) Enhancing code generation via self-debugging, where models iteratively refine solutions by generating and analyzing test cases~\cite{Chen2022CodeTCG, Zhang2023ALGOSA, Shinn2023ReflexionLA, Jiao2024PreferenceOF, Zeng2025ACECODERAC}. 
(2) Improving code evaluation by leveraging LLMs to generate diverse test cases, as used in recent code generation evaluation benchmarks~\cite{Liu2023IsYC, Du2024MercuryAC, Yu2024HumanEvalPA, Jain2024LiveCodeBenchHA}.
Despite these advancements, a systematic study on the standalone capability of LLMs in test-case generation
remains an open challenge. 
A closely related benchmark for algorithmic problem test-case generation is TestEval~\cite{Wang2024TESTEVALBL}, which evaluates test generation for LeetCode problems but relies on traditional Line and Branch Coverage assessment (illustrated in Appendix~\ref{app:related-work}), which may be inadequate for algorithmic problem settings.
For instance, the 6.7B DeepSeek-Coder achieves over 90\% in TestEval, with top models nearing 100\%.
Our work shifts the focus to more challenging CodeForces competition problems and introduces two novel and challenging tasks. Even state-of-the-art models achieve only around 40\% on the task of Fault Exposure.

\section{\ours Benchmark}
This section discusses the \ours benchmark construction and task settings, as illustrated in \autoref{fig:main-example}.

\subsection{Source Data Collection}
We first outline the data collection process, which involves collecting \emph{problems} with corresponding correct and incorrect \emph{human-written solutions}.

\paragraph{Problem Collection.}

We collect algorithmic problems from \cf contests held between January 1, 2024, and December 30, 2024. This time frame falls outside the pretraining period of most existing foundation models, reducing potential data memorization concerns.
Our goal is to curate a dataset that (1) includes a substantial number of incorrect submissions and (2) ensures accurate offline evaluation. 
To achieve this, we apply a series of filtering steps. 
First, we exclude problems requiring special judge functionalities (detailed in Appendix~\ref{app:judge}), as these allow multiple valid outputs for a single input, often leading to unreliable evaluations. 
Next, we verify each problem by running ten correct human-written solutions sourced from \cf, ensuring they consistently produce identical outputs for the same test inputs. 
Additionally, we remove problems with fewer than 1,000 online incorrect submissions to ensure a diverse range of mistakes.
After this selection process, our final dataset includes a total of \nproblem problems. 
\autoref{fig:problem-difficulty-distribution} in Appendix presents the distribution of problem difficulty ratings.

\paragraph{Human-written Solution Collection.}
For each problem in \ours, we collect 200 incorrect submissions (\ie human-written solutions along with their evaluation outcomes) from the \cf platform. 
The platform provides detailed error types for each incorrect submission, such as ``Memory Limit Exceeded'', ``Time Limit Exceeded'', ``Runtime Error'', and ``Wrong Answer''. 
Additionally, it specifies the test case index where the error occurred (\eg ``Wrong answer on test 5''), with higher indices generally indicating more complex or inherent errors.
Leveraging this information, we categorize incorrect submissions into three difficulty levels: \emph{Easy}, \emph{Medium}, and \emph{Hard}.
(Detailed difficulty definition and distribution are illustrated in Appendix~\ref{app:human-written-solutions})
To ensure diversity, we select code written in three widely used programming languages: C++, Python, and Java. We begin by crawling the submission logs for each problem, which contain the complete history of contestant submissions. These logs are carefully filtered based on test case failures and programming language criteria, resulting in a comprehensive dataset of \nsubmission submissions.
(Detailed dataset collection and sampling pipeline are illustrated in Appendix~\ref{app:human-written-solutions})

We next outline the construction process for the two evaluation tasks in \ours.

\subsection{Fault Coverage Evaluation (Task 1)}
This task assesses the LLM's ability to generate comprehensive test inputs that effectively detect faulty code implementations. 
Specifically, given the description of an algorithmic problem, the LLM must thoroughly understand the problem and generate a specified number of test cases that maximize coverage of incorrect solution scenarios.
Let \(\mathcal{T_N} = \{t_1, t_2, \dots, t_N\}\) represent the set of \(N\) test inputs generated by the LLM. For each test input \(t_i\), let \(\mathcal{F}(t_i)\) denote the subset of incorrect submissions it detects, drawn from the complete set of incorrect solutions \(\mathcal{F}_{\text{total}}\),
the final score for this task is defined as the coverage rate of incorrect solutions when generating \(N\) test inputs:

\begin{small}
\[
Cov@N = \frac{\left|\bigcup_{i=1}^{N} \mathcal{F}(t_i)\right|}{|\mathcal{F}_{\text{total}}|}
\]
\end{small}

\noindent It quantifies the LLM’s effectiveness in generating diverse and impactful test cases that expose incorrect implementations.

\subsection{Fault Exposure Evaluation (Task 2)}
This task is inspired by the hacking phase in CodeForces competitions, where participants analyze others' solutions and attempt to ``hack'' them by providing inputs that reveal flaws in the code.  
The goal is to assess the LLM's ability to understand both the problem and the specific errors present in the faulty implementation. 
Given the description of an algorithmic problem and a single faulty code implementation \( f_i \) within the sampled set \( \mathcal{F} \) (a strategically sampled subset of \( \mathcal{F}_{\text{total}} \)), Task 2 requires the LLM to generate a single test input \( t_i \) to exploit the fault.
The \emph{Fault Exposure Rate} is computed as:
\[
~\text{Fault Exposure Rate} = \frac{1}{|\mathcal{F}|} \sum_{f_i \in \mathcal{F}} e(f_i, t_i) \text{, where}
\]
\begin{small}
\[
e(f_i, t_i) = \begin{cases} 
1, & \text{if } t_i \text{ successfully exposes fault in } f_i, \\[1mm]
0, & \text{otherwise.}
\end{cases}
\]
\end{small}

\noindent It measures both general test case generation capabilities and targeted fault detection performance.

\section{Experiment}
\subsection{Experiment Setup}
We evaluate 11 series of open-source models, including 
Qwen2.5~\cite{Yang2024Qwen25TR} and Qwen2.5-Coder~\cite{Hui2024Qwen25CoderTR}, Qwen3~\cite{qwen3}, QwQ~\cite{qwq32b}, Llama-3.1\&3.3~\cite{Dubey2024TheL3}, Gemma-3~\cite{gemmateam2025gemma3technicalreport}, DeepSeek-R1~\cite{DeepSeekAI2025DeepSeekR1IR}, Mistral-Small~\cite{Jiang2023Mistral7}, Codestral~\cite{Codestral}, and SeedCoder~\cite{seedcoder}.
We also evaluate two series of proprietary models, including 
GPT-4o~\cite{openai2024gpt4o} and GPT-4.1~\cite{openai2025gpt41}.
Appendix \ref{app:model_info} details the parameter settings and model configurations.
We evaluate the models with both \textbf{Direct Output} and \textbf{Chain-of-Thought} prompts, with prompting examples presented in Appendix \ref{app:prompt}.
We utilize the \textbf{sandbox environment} from ExecEval~\cite{Khan2023xCodeEvalAL} for code execution and test input evaluation, ensuring secure execution and reliable assessment of results.
To approximate \textbf{human-expert-level performance} on \ours, we randomly sampled 20 problems from the dataset. Two human experts, with Codeforces ratings of 2080 and 2237, independently completed both Task 1 and Task 2 for each problem. Their performance was then averaged to obtain the final assessment.

\begin{figure*}[!t]
    \begin{minipage}[t]{0.55\textwidth}
        \vspace{0pt}
        \centering
        \renewcommand\arraystretch{1.15}
        \addtolength{\tabcolsep}{-0.27em}
        \resizebox{\linewidth}{!}{
        \footnotesize
        \begin{tabular}{l *{4}{>{\centering\arraybackslash}p{0.6cm}}c*{4}{>{\centering\arraybackslash}p{0.55cm}}}
        \toprule
        \multirow{2}{*}{\textbf{Models}} & \multicolumn{4}{c}{\textbf{T1: Fault Coverage}} & & \multicolumn{4}{c}{\textbf{T2: Fault Exposure}} \\
        \cmidrule{2-5}\cmidrule{7-10}
         & c@1 & c@5 & c@10 & c@20 &  & Easy & Med. & Hard & \textbf{Ovr.} \\
        \midrule
        Human Expert & 56.2 & 85.7 & 93.5 & 97.2 &  & 95.0 & 92.5 & 91.8 & 93.3 \\
        \midrule
        GPT-4.1 & 45.3 & 67.5 & 74.1 & 80.0 &  & 42.9 & 34.3 & 30.3 & 36.5 \\
        GPT-4.1-mini & 38.8 & 63.2 & 68.5 & 72.6 &  & 39.2 & 32.4 & 27.4 & 33.6 \\
        GPT-4o & 36.4 & 60.3 & 69.7 & 73.5 &  & 37.5 & 30.5 & 25.2 & 31.7 \\
        \noalign{\vskip 0.5ex}\hdashline\noalign{\vskip 0.5ex}
        Qwen3-8B & \cellcolor{red!5}46.2 & \cellcolor{red!20}78.5 & \cellcolor{red!20}87.9 & \cellcolor{red!20}92.1 &  & 48.6 & \cellcolor{red!20}39.8 & 33.1 & \cellcolor{red!5}41.3 \\
        Qwen3-32B & \cellcolor{red!35}50.8 & \cellcolor{red!35}82.3 & \cellcolor{red!35}92.6 & \cellcolor{red!35}95.7 &  & \cellcolor{red!35}52.7 & \cellcolor{red!35}42.5 & \cellcolor{red!5}33.2 & \cellcolor{red!35}43.8 \\
        R1-Distill-Qwen-32B & 31.9 & 65.3 & 75.6 & 82.6 &  & \cellcolor{red!5}48.7 & \cellcolor{red!5}39.7 & \cellcolor{red!35}33.9 & \cellcolor{red!20}41.6 \\
        QwQ-32B & 37.3 & 58.9 & 67.6 & 78.3 &  & \cellcolor{red!20}49.4 & 38.0 & 30.2 & 40.2 \\
        Qwen2.5-7B & 38.6 & 65.4 & 73.0 & 79.1 &  & 39.8 & 34.2 & 29.5 & 35.0 \\
        Qwen2.5-Coder-7B & 36.7 & 63.2 & 70.5 & 76.5 &  & 37.2 & 33.1 & 29.9 & 33.7 \\
        Qwen2.5-32B & 44.4 & 70.9 & 79.6 & 88.4 &  & 38.8 & 30.4 & 25.5 & 32.3 \\
        Qwen2.5-Coder-32B & 35.8 & 66.7 & 81.8 & 89.7 &  & 40.5 & 34.0 & 27.3 & 34.6 \\
        Qwen2.5-72B & 38.2 & 57.8 & 65.2 & 73.1 &  & 33.6 & 27.5 & 24.2 & 29.0 \\
        Llama-3.1-70B & \cellcolor{red!20}47.8 & \cellcolor{red!5}75.4 & 84.8 & \cellcolor{red!5}90.9 &  & 37.9 & 33.5 & 30.5 & 34.3 \\
        Llama-3.3-70B & 43.2 & 72.5 & 81.2 & 88.6 &  & 33.8 & 27.9 & 25.2 & 29.5 \\
        Mistral-Small-24B & 35.5 & 71.9 & 80.4 & 88.3 &  & 37.4 & 31.9 & 28.4 & 33.1 \\
        Codestral-22B & 34.8 & 68.8 & \cellcolor{red!5}87.4 & 90.8 &  & 34.9 & 28.2 & 26.4 & 30.3 \\
        Gemma-3-12B & 30.4 & 54.6 & 61.0 & 65.3 &  & 35.7 & 31.9 & \cellcolor{red!20}33.3 & 33.8 \\
        Gemma-3-27B & 32.4 & 55.6 & 64.1 & 70.7 &  & 34.3 & 28.3 & 28.3 & 30.7 \\
        Seed-Coder-8B & 30.7 & 63.2 & 75.6 & 87.4 &  & 34.5 & 28.1 & 25.5 & 29.9 \\
        \bottomrule
        \end{tabular}
        }
        \captionof{table}{Performance of the evaluated LLMs with CoT reasoning on \ours. For Task 1, we report Coverage@N; for Task 2, we report fault exposure rate.}
        \label{tab:main_results}
    \end{minipage}%
    \hfill
    \begin{minipage}[t]{0.43\textwidth}
        \vspace{0pt}
        \centering
        \includegraphics[width=1\linewidth]{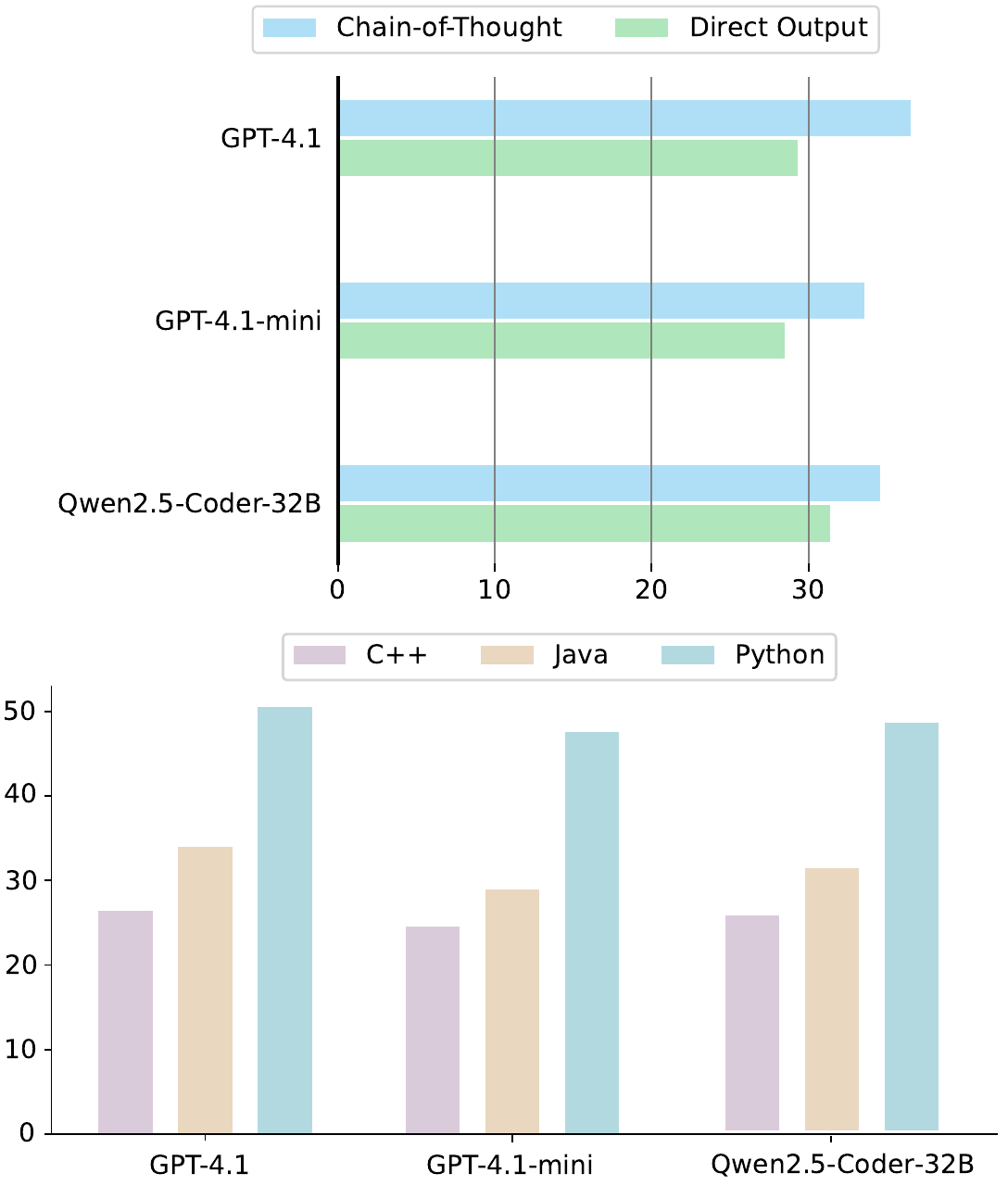}
        \captionof{figure}{(Top) Performance comparison between CoT prompting and direct-output prompting for Task 2. (Bottom) Overall model performance using CoT prompting across C++, Java, and Python in Task 2.}
        \label{fig:analysis}
    \end{minipage}

\end{figure*}

\subsection{Experimental Results and Analysis}
\hyperref[tab:main_results]{Table~\ref*{tab:main_results}} illustrates the model performance on \ours. Our key findings are as follows:

\paragraph{\ours presents substantial challenges for current models.}
The \ours benchmark is highly challenging, as evidenced by the significant performance gap between models and human experts on both tasks. This gap is particularly pronounced in Task 2 (Fault Exposure), where human experts achieve a 93.3\% fault exposure rate, more than double the best-performing model, Qwen3-32B (43.8\%). While Task 1 (Fault Coverage) also shows a considerable gap, models achieve relatively higher scores, suggesting that generating a broad set of test cases is more tractable than triggering specific code flaws. Furthermore, we observe that Task 2 yields more stable and reproducible results across multiple evaluation runs, whereas Task 1 scores exhibit higher variance, likely due to the stochastic nature of generating a diverse set of test inputs.

\paragraph{Open-source models compete with or surpass proprietary counterparts.}
Our results indicate that leading open-source models are highly competitive. In Task 1 (Fault Coverage), several open-source models, including Qwen3-32B (50.8 cov@1) and Llama-3.1-70B (47.8 cov@1), outperform the best proprietary model, GPT-4.1 (45.3 cov@1). The trend continues at higher N values, where Qwen3-32B's cov@20 score of 95.7 significantly surpasses GPT-4.1's 80.0. In the more reasoning-intensive Task 2, while the Qwen3 series leads, GPT-4.1 shows strong performance with a 36.5\% fault exposure rate, outperforming all other general-purpose open-source models like Llama-3.1-70B (34.3\%). This highlights a competitive landscape where proprietary models do not hold a universal advantage in our benchmark.

\paragraph{Reasoning LLMs outperform general-purpose LLMs on both tasks.}
Reasoning-oriented models, such as the Qwen3 series, demonstrate superior performance on both tasks. Notably, Qwen3-32B achieves the highest scores in Task 1 across all metrics (e.g., 50.8 cov@1 and 95.7 cov@20), clearly surpassing strong general-purpose models such as Llama-3.1-70B, as well as proprietary models like GPT-4.1. This performance gap becomes even more pronounced in Task 2, which demands deeper analytical capabilities. Qwen3-32B and R1-Distill-Qwen-32B attain the top two fault exposure rates, at 43.8\% and 41.6\% respectively, with a substantial margin over all other evaluated models. These results suggest that reasoning models excel because they are better equipped to analyze algorithmic problem descriptions, systematically identify possible fault patterns, and generate high-quality test inputs.

\paragraph{CoT prompts vs direct-output prompts.}  
Our experiments reveal that CoT prompting significantly outperforms direct-output prompting in generating test cases (\autoref{fig:analysis}). This advantage stems from CoT’s structured reasoning process, which guides the model through intermediate steps before arriving at the final output. Such an approach is particularly beneficial for the complex tasks in \ours, where systematic thinking is crucial. When generating test cases for fault exposure, CoT prompting led to more effective fault detection, especially in challenging edge cases. This suggests that models benefit from explicit reasoning steps, as they help decompose intricate problems and improve fault exposure rate in nuanced scenarios.

\paragraph{Comparison of fault exposure results across different programming languages.}  
In Fault Exposure task, model performance varies across programming languages. As shown in \autoref{fig:analysis}, models generally achieve higher fault exposure rates on Python solutions, likely due to Python’s dynamic typing, flexible syntax, and interpreted execution, which facilitate the generation of diverse test cases that reveal faults. 
In contrast, C++ and Java exhibit lower fault exposure rates, possibly due to their strict type enforcement, manual memory management, and compiled execution, which can limit the likelihood of generating inputs that reveal subtle faults. 
Addressing these differences through targeted adaptation could help improve fault detection across a wider range of programming languages.

\paragraph{Performance breakdown to four major error types.}
\autoref{tab:task2_results} provides a detailed breakdown of model performance on task 2 across four major error types: Wrong Answer (WA), Runtime Error (RE), Time Limit Exceeded (TLE), and Memory Limit Exceeded (MLE). A clear trend emerges from the data: models generally demonstrate stronger capabilities in detecting logical and execution-related faults (WA and RE) compared to resource-based faults (TLE and MLE). This suggests that current LLMs are generally better at detecting logical or edge-case errors than time or memory inefficiencies.

Notably, the best-performing models, Qwen3-32B and R1-Distill-Qwen-32B, exhibit significantly higher accuracy on Wrong Answer type. Given that WA constitutes the largest proportion of error types, their superior performance in this category is the primary driver of their high overall accuracy (43.8\% and 41.6\%, respectively). This finding underscores that the strength of these advanced reasoning models lies in their enhanced ability to construct precise test cases that target and expose logical inconsistencies within code.

\begin{table}[!t]
    \centering
    \renewcommand\arraystretch{1.15}
    \addtolength{\tabcolsep}{-0.27em}
    \resizebox{\linewidth}{!}{
        \small
        \begin{tabular}{lccccc}
            \toprule
            \textbf{Models} & \textbf{WA} & \textbf{RE} & \textbf{TLE} & \textbf{MLE} & \textbf{Ovr.} \\
            \midrule
            GPT-4.1 & 42.0 & 35.4 & 20.9 & 25.1 & 36.5 \\
            GPT-4.1-mini & 39.3 & 32.0 & 17.4 & 24.6 & 33.6 \\
            GPT-4o & 37.1 & 29.0 & 16.4 & 25.1 & 31.7 \\
            \noalign{\vskip 0.5ex}\hdashline\noalign{\vskip 0.5ex}
            Qwen3-8B & 48.0 & 39.0 & 22.8 & 26.9 & 41.3 \\
            Qwen3-32B & 52.2 & 38.7 & 21.2 & 22.3 & 43.8 \\
            R1-Distill-Qwen-32B & 48.0 & 37.8 & 23.9 & 30.3 & 41.6 \\
            QwQ-32B & 49.1 & 35.1 & 16.4 & 14.9 & 40.2 \\
            Qwen2.5-7B & 38.7 & 37.4 & 23.6 & 31.4 & 35.0 \\
            Qwen2.5-Coder-7B & 35.5 & 39.8 & 26.5 & 36.0 & 33.7 \\
            Qwen2.5-32B & 36.8 & 34.7 & 18.0 & 28.6 & 32.3 \\
            Qwen2.5-Coder-32B & 38.9 & 38.3 & 21.0 & 26.9 & 34.6 \\
            Qwen2.5-72B & 33.0 & 30.5 & 16.8 & 20.6 & 29.0 \\
            Llama-3.1-70B & 36.5 & 33.9 & 27.7 & 36.6 & 34.3 \\
            Llama-3.3-70B & 32.9 & 30.2 & 19.3 & 24.6 & 29.5 \\
            Mistral-Small-24B & 35.8 & 37.4 & 23.1 & 40.0 & 33.1 \\
            Codestral-22B & 33.1 & 35.0 & 20.3 & 31.4 & 30.3 \\
            Gemma-3-12B & 35.8 & 35.1 & 27.7 & 30.9 & 33.8 \\
            Gemma-3-27B & 33.5 & 33.3 & 22.3 & 21.7 & 30.7 \\
            Seed-Coder-8B & 32.8 & 31.7 & 20.5 & 31.4 & 29.9 \\
            \bottomrule
        \end{tabular}
    }
    \caption{Performance breakdown of evaluated LLMs on task 2 (fault exposure), reported by four error types.}
    \label{tab:task2_results}
\end{table}

\section{Conclusion}
We introduce \ours, a new benchmark for evaluating LLMs in test-case generation for algorithmic problems, with a focus on Fault Coverage and Fault Exposure.
Our results show that all the evaluated LLMs struggle with harder faults, highlighting the challenge of automated test-case generation. 
Although CoT prompting enhances performance, a substantial gap remains between frontier LLMs and human experts.
These findings emphasize the need for further research to enhance LLMs' capabilities in generating high-quality test cases and their practical application in software testing.

\section*{Limitations}
While \ours provides a comprehensive benchmark for evaluating LLMs in test-case generation, several limitations warrant further investigation:
(1) Our evaluation primarily emphasizes quantitative performance indicators, such as fault coverage and exposure rates, which might not capture the nuanced failure modes that may arise in LLM-generated test cases. Future work could include more detailed error analyses to uncover specific failure patterns and model weaknesses.
(2) The difficulty levels (Easy, Medium, Hard) in \ours are determined by the test case index where an incorrect solution first fails. While this provides a reasonable estimate of error complexity, it does not explicitly categorize the types of mistakes.
(3) Our benchmark focuses on correctness-based faults and does not systematically test performance bottlenecks (\eg time limit exceeded, memory limit exceeded). Although some incorrect submissions fail due to resource constraints, we do not explicitly assess whether LLMs generate test cases that effectively expose computational complexity flaws.
(4) While \ours targets test-case generation, practical software testing also requires identifying the root cause and location of bugs. Future work could extend this to more holistic debugging and fault localization tasks.

\bibliography{anthology,custom, llm}

\begin{thebibliography}{31}
\expandafter\ifx\csname natexlab\endcsname\relax\def\natexlab#1{#1}\fi

\bibitem[{Austin et~al.(2021)Austin, Odena, Nye, Bosma, Michalewski, Dohan, Jiang, Cai, Terry, Le, and Sutton}]{Austin2021ProgramSW}
Jacob Austin, Augustus Odena, Maxwell Nye, Maarten Bosma, Henryk Michalewski, David Dohan, Ellen Jiang, Carrie~J. Cai, Michael Terry, Quoc~V. Le, and Charles Sutton. 2021.
\newblock \href {https://api.semanticscholar.org/CorpusID:237142385} {Program synthesis with large language models}.
\newblock \emph{ArXiv}, abs/2108.07732.

\bibitem[{Chen et~al.(2022)Chen, Zhang, Nguyen, Zan, Lin, Lou, and Chen}]{Chen2022CodeTCG}
Bei Chen, Fengji Zhang, A.~Nguyen, Daoguang Zan, Zeqi Lin, Jian-Guang Lou, and Weizhu Chen. 2022.
\newblock \href {https://api.semanticscholar.org/CorpusID:250920542} {Codet: Code generation with generated tests}.
\newblock \emph{ArXiv}, abs/2207.10397.

\bibitem[{DeepSeek-AI et~al.(2025)DeepSeek-AI, Guo, Yang, Zhang, Song, Zhang, Xu, Zhu, Ma, Wang, Bi, Zhang, Yu, Wu, Wu, Gou, Shao, Li, Gao, Liu, Xue, Wang, Wu, Feng, Lu, Zhao, Deng, Zhang, Ruan, Dai, Chen, Ji, Li, Lin, Dai, Luo, Hao, Chen, Li, Zhang, Bao, Xu, Wang, Ding, Xin, Gao, Qu, Li, Guo, Li, Wang, Chen, Yuan, Qiu, Li, Cai, Ni, Liang, Chen, Dong, Hu, Gao, Guan, Huang, Yu, Wang, Zhang, Zhao, Wang, Zhang, Xu, Xia, Zhang, Zhang, Tang, Li, Wang, Li, Tian, Huang, Zhang, Wang, Chen, Du, Ge, Zhang, Pan, Wang, Chen, Jin, Chen, Lu, Zhou, Chen, Ye, Wang, Yu, Zhou, Pan, Li, Zhou, Wu, Yun, Pei, Sun, Wang, Zeng, Zhao, Liu, Liang, Gao, Yu, Zhang, Xiao, An, Liu, Wang, Chen, Nie, Cheng, Liu, Xie, Liu, Yang, Li, Su, Lin, Li, Jin, Shen, Chen, Sun, Wang, Song, Zhou, Wang, Shan, Li, Wang, Wei, Zhang, Xu, Li, Zhao, Sun, Wang, Yu, Zhang, Shi, Xiong, He, Piao, Wang, Tan, Ma, Liu, Guo, Ou, Wang, Gong, Zou, He, Xiong, Luo, mei You, Liu, Zhou, Zhu, Huang, Li, Zheng, Zhu, Ma, Tang, Zha, Yan, Ren, Ren, Sha, Fu, Xu, Xie, guo Zhang,
  Hao, Ma, Yan, Wu, Gu, Zhu, Liu, Li, Xie, Song, Pan, Huang, Xu, Zhang, and Zhang}]{DeepSeekAI2025DeepSeekR1IR}
DeepSeek-AI, Daya Guo, Dejian Yang, Haowei Zhang, Jun-Mei Song, Ruoyu Zhang, Runxin Xu, Qihao Zhu, Shirong Ma, Peiyi Wang, Xiaoling Bi, Xiaokang Zhang, Xingkai Yu, Yu~Wu, Z.~F. Wu, Zhibin Gou, Zhihong Shao, Zhuoshu Li, Ziyi Gao, Aixin Liu, Bing Xue, Bing-Li Wang, Bochao Wu, Bei Feng, Chengda Lu, Chenggang Zhao, Chengqi Deng, Chenyu Zhang, Chong Ruan, Damai Dai, Deli Chen, Dong-Li Ji, Erhang Li, Fangyun Lin, Fucong Dai, Fuli Luo, Guangbo Hao, Guanting Chen, Guowei Li, H.~Zhang, Han Bao, Hanwei Xu, Haocheng Wang, Honghui Ding, Huajian Xin, Huazuo Gao, Hui Qu, Hui Li, Jianzhong Guo, Jiashi Li, Jiawei Wang, Jingchang Chen, Jingyang Yuan, Junjie Qiu, Junlong Li, Jiong Cai, Jiaqi Ni, Jian Liang, Jin Chen, Kai Dong, Kai Hu, Kaige Gao, Kang Guan, Kexin Huang, Kuai Yu, Lean Wang, Lecong Zhang, Liang Zhao, Litong Wang, Liyue Zhang, Lei Xu, Leyi Xia, Mingchuan Zhang, Minghua Zhang, M.~Tang, Meng Li, Miaojun Wang, Mingming Li, Ning Tian, Panpan Huang, Peng Zhang, Qiancheng Wang, Qinyu Chen, Qiushi Du, Ruiqi Ge, Ruisong
  Zhang, Ruizhe Pan, Runji Wang, R.~J. Chen, R.~L. Jin, Ruyi Chen, Shanghao Lu, Shangyan Zhou, Shanhuang Chen, Shengfeng Ye, Shiyu Wang, Shuiping Yu, Shunfeng Zhou, Shuting Pan, S.~S. Li, Shuang Zhou, Shao-Kang Wu, Tao Yun, Tian Pei, Tianyu Sun, T.~Wang, Wangding Zeng, Wanjia Zhao, Wen Liu, Wenfeng Liang, Wenjun Gao, Wen-Xia Yu, Wentao Zhang, W.~L. Xiao, Wei An, Xiaodong Liu, Xiaohan Wang, Xiaokang Chen, Xiaotao Nie, Xin Cheng, Xin Liu, Xin Xie, Xingchao Liu, Xinyu Yang, Xinyuan Li, Xuecheng Su, Xuheng Lin, X.~Q. Li, Xiangyu Jin, Xi-Cheng Shen, Xiaosha Chen, Xiaowen Sun, Xiaoxiang Wang, Xinnan Song, Xinyi Zhou, Xianzu Wang, Xinxia Shan, Y.~K. Li, Y.~Q. Wang, Y.~X. Wei, Yang Zhang, Yanhong Xu, Yao Li, Yao Zhao, Yaofeng Sun, Yaohui Wang, Yi~Yu, Yichao Zhang, Yifan Shi, Yi~Xiong, Ying He, Yishi Piao, Yisong Wang, Yixuan Tan, Yiyang Ma, Yiyuan Liu, Yongqiang Guo, Yuan Ou, Yuduan Wang, Yue Gong, Yu-Jing Zou, Yujia He, Yunfan Xiong, Yu-Wei Luo, Yu~mei You, Yuxuan Liu, Yuyang Zhou, Y.~X. Zhu, Yanping Huang, Yao Li,
  Yi~Zheng, Yuchen Zhu, Yunxiang Ma, Ying Tang, Yukun Zha, Yuting Yan, Zehui Ren, Zehui Ren, Zhangli Sha, Zhe Fu, Zhean Xu, Zhenda Xie, Zhen guo Zhang, Zhewen Hao, Zhicheng Ma, Zhigang Yan, Zhiyu Wu, Zihui Gu, Zijia Zhu, Zijun Liu, Zi-An Li, Ziwei Xie, Ziyang Song, Zizheng Pan, Zhen Huang, Zhipeng Xu, Zhongyu Zhang, and Zhen Zhang. 2025.
\newblock \href {https://api.semanticscholar.org/CorpusID:275789950} {Deepseek-r1: Incentivizing reasoning capability in llms via reinforcement learning}.

\bibitem[{Du et~al.(2024)Du, Luu, Ji, Liu, and Ng}]{Du2024MercuryAC}
Mingzhe Du, Anh~Tuan Luu, Bin Ji, Qian Liu, and See-Kiong Ng. 2024.
\newblock \href {https://api.semanticscholar.org/CorpusID:269758190} {Mercury: A code efficiency benchmark for code large language models}.
\newblock In \emph{Neural Information Processing Systems}.

\bibitem[{El-Kishky et~al.(2025)El-Kishky, Wei, Saraiva, Minaev, Selsam, Dohan, Song, Lightman, Clavera, Pachocki, Tworek, Kuhn, Kaiser, Chen, Schwarzer, Rohaninejad, McAleese, o3~contributors, Murk, Garg, Shu, Sidor, Kosaraju, and Zhou}]{ElKishky2025CompetitivePW}
OpenAI~Ahmed El-Kishky, Alexander Wei, Andre Saraiva, Borys Minaev, Daniel Selsam, David Dohan, Francis Song, Hunter Lightman, Ignasi Clavera, Jakub~W. Pachocki, Jerry Tworek, Lorenz Kuhn, Lukasz Kaiser, Mark Chen, Max Schwarzer, Mostafa Rohaninejad, Nat McAleese, o3~contributors, Oleg Murk, Rhythm Garg, Rui Shu, Szymon Sidor, Vineet Kosaraju, and Wenda Zhou. 2025.
\newblock \href {https://api.semanticscholar.org/CorpusID:276258630} {Competitive programming with large reasoning models}.

\bibitem[{Hendrycks et~al.(2021)Hendrycks, Basart, Kadavath, Mazeika, Arora, Guo, Burns, Puranik, He, Song, and Steinhardt}]{Hendrycks2021MeasuringCC}
Dan Hendrycks, Steven Basart, Saurav Kadavath, Mantas Mazeika, Akul Arora, Ethan Guo, Collin Burns, Samir Puranik, Horace He, Dawn~Xiaodong Song, and Jacob Steinhardt. 2021.
\newblock \href {https://api.semanticscholar.org/CorpusID:234790100} {Measuring coding challenge competence with apps}.
\newblock \emph{ArXiv}, abs/2105.09938.

\bibitem[{Huang et~al.(2023)Huang, Vora, Liang, and Leskovec}]{Huang2023MLAgentBenchEL}
Qian Huang, Jian Vora, Percy Liang, and Jure Leskovec. 2023.
\newblock \href {https://api.semanticscholar.org/CorpusID:263671541} {Mlagentbench: Evaluating language agents on machine learning experimentation}.
\newblock In \emph{International Conference on Machine Learning}.

\bibitem[{Hui et~al.(2024)Hui, Yang, Cui, Yang, Liu, Zhang, Liu, Zhang, Yu, Dang, Yang, Men, Huang, Quan, Ren, Ren, Zhou, and Lin}]{Hui2024Qwen25CoderTR}
Binyuan Hui, Jian Yang, Zeyu Cui, Jiaxi Yang, Dayiheng Liu, Lei Zhang, Tianyu Liu, Jiajun Zhang, Bowen Yu, Kai Dang, An~Yang, Rui Men, Fei Huang, Shanghaoran Quan, Xingzhang Ren, Xuancheng Ren, Jingren Zhou, and Junyang Lin. 2024.
\newblock \href {https://api.semanticscholar.org/CorpusID:272707390} {Qwen2.5-coder technical report}.
\newblock \emph{ArXiv}, abs/2409.12186.

\bibitem[{Jain et~al.(2024)Jain, Han, Gu, Li, Yan, Zhang, Wang, Solar-Lezama, Sen, and Stoica}]{Jain2024LiveCodeBenchHA}
Naman Jain, King Han, Alex Gu, Wen-Ding Li, Fanjia Yan, Tianjun Zhang, Sida~I. Wang, Armando Solar-Lezama, Koushik Sen, and Ion Stoica. 2024.
\newblock \href {https://api.semanticscholar.org/CorpusID:268379413} {Livecodebench: Holistic and contamination free evaluation of large language models for code}.
\newblock \emph{ArXiv}, abs/2403.07974.

\bibitem[{Jiang et~al.(2023)Jiang, Sablayrolles, Mensch, Bamford, Chaplot, de~Las~Casas, Bressand, Lengyel, Lample, Saulnier, Lavaud, Lachaux, Stock, Scao, Lavril, Wang, Lacroix, and Sayed}]{Jiang2023Mistral7}
Albert~Qiaochu Jiang, Alexandre Sablayrolles, Arthur Mensch, Chris Bamford, Devendra~Singh Chaplot, Diego de~Las~Casas, Florian Bressand, Gianna Lengyel, Guillaume Lample, Lucile Saulnier, L'elio~Renard Lavaud, Marie-Anne Lachaux, Pierre Stock, Teven~Le Scao, Thibaut Lavril, Thomas Wang, Timoth{\'e}e Lacroix, and William~El Sayed. 2023.
\newblock \href {https://api.semanticscholar.org/CorpusID:263830494} {Mistral 7b}.
\newblock \emph{ArXiv}, abs/2310.06825.

\bibitem[{Jiao et~al.(2024)Jiao, Guo, Zhang, Chen, Joty, and Wei}]{Jiao2024PreferenceOF}
Fangkai Jiao, Geyang Guo, Xingxing Zhang, Nancy~F. Chen, Shafiq Joty, and Furu Wei. 2024.
\newblock \href {https://api.semanticscholar.org/CorpusID:274233865} {Preference optimization for reasoning with pseudo feedback}.
\newblock \emph{ArXiv}, abs/2411.16345.

\bibitem[{Jimenez et~al.(2023)Jimenez, Yang, Wettig, Yao, Pei, Press, and Narasimhan}]{Jimenez2023SWEbenchCL}
Carlos~E. Jimenez, John Yang, Alexander Wettig, Shunyu Yao, Kexin Pei, Ofir Press, and Karthik Narasimhan. 2023.
\newblock \href {https://api.semanticscholar.org/CorpusID:263829697} {Swe-bench: Can language models resolve real-world github issues?}
\newblock \emph{ArXiv}, abs/2310.06770.

\bibitem[{Khan et~al.(2023)Khan, Bari, Long, Wang, Parvez, and Joty}]{Khan2023xCodeEvalAL}
Mohammad Abdullah~Matin Khan, M~Saiful Bari, Do~Xuan Long, Weishi Wang, Md.~Rizwan Parvez, and Shafiq~R. Joty. 2023.
\newblock \href {https://api.semanticscholar.org/CorpusID:257365592} {xcodeeval: A large scale multilingual multitask benchmark for code understanding, generation, translation and retrieval}.
\newblock \emph{ArXiv}, abs/2303.03004.

\bibitem[{Li et~al.(2022)Li, Choi, Chung, Kushman, Schrittwieser, Leblond, Tom, Eccles, Keeling, Gimeno, Lago, Hubert, Choy, de, d’Autume, Babuschkin, Chen, Huang, Welbl, Gowal, Alexey, Cherepanov, Molloy, Mankowitz, Robson, Kohli, de, Freitas, Kavukcuoglu, and Vinyals}]{Li2022CompetitionlevelCG}
Yujia Li, David Choi, Junyoung Chung, Nate Kushman, Julian Schrittwieser, R{\'e}mi Leblond, Tom, Eccles, James Keeling, Felix Gimeno, Agustin~Dal Lago, Thomas Hubert, Peter Choy, Cyprien de, Masson d’Autume, Igor Babuschkin, Xinyun Chen, Po-Sen Huang, Johannes Welbl, Sven Gowal, Alexey, Cherepanov, James Molloy, Daniel~Jaymin Mankowitz, Esme~Sutherland Robson, Pushmeet Kohli, Nando de, Freitas, Koray Kavukcuoglu, and Oriol Vinyals. 2022.
\newblock \href {https://api.semanticscholar.org/CorpusID:246527904} {Competition-level code generation with alphacode}.
\newblock \emph{Science}, 378:1092 -- 1097.

\bibitem[{Liu et~al.(2023)Liu, Xia, Wang, and Zhang}]{Liu2023IsYC}
Jiawei Liu, Chun Xia, Yuyao Wang, and Lingming Zhang. 2023.
\newblock \href {https://api.semanticscholar.org/CorpusID:258437095} {Is your code generated by chatgpt really correct? rigorous evaluation of large language models for code generation}.
\newblock \emph{ArXiv}, abs/2305.01210.

\bibitem[{Meta(2024)}]{Dubey2024TheL3}
Meta. 2024.
\newblock \href {https://api.semanticscholar.org/CorpusID:271571434} {The llama 3 herd of models}.
\newblock \emph{ArXiv}, abs/2407.21783.

\bibitem[{OpenAI(2024)}]{openai2024gpt4o}
OpenAI. 2024.
\newblock \href {https://openai.com/index/hello-gpt-4o/} {Hello gpt-4o}.

\bibitem[{OpenAI(2025)}]{openai2025gpt41}
OpenAI. 2025.
\newblock \href {https://openai.com/index/gpt-4-1/} {Introducing gpt-4.1 in the api}.

\bibitem[{Seed(2025)}]{seedcoder}
ByteDance Seed. 2025.
\newblock \href {https://github.com/ByteDance-Seed/Seed-Coder/blob/master/Seed-Coder.pdf} {Seed-coder: Let the code model curate data for itself}.

\bibitem[{Shinn et~al.(2023)Shinn, Cassano, Labash, Gopinath, Narasimhan, and Yao}]{Shinn2023ReflexionLA}
Noah Shinn, Federico Cassano, Beck Labash, Ashwin Gopinath, Karthik Narasimhan, and Shunyu Yao. 2023.
\newblock \href {https://api.semanticscholar.org/CorpusID:258833055} {Reflexion: language agents with verbal reinforcement learning}.
\newblock In \emph{Neural Information Processing Systems}.

\bibitem[{Team et~al.(2025)Team, Kamath, Ferret, Pathak, Vieillard, Merhej, Perrin, Matejovicova, Ramé, Rivière, Rouillard, Mesnard, Cideron, bastien Grill, Ramos, Yvinec, Casbon, Pot, Penchev, Liu, Visin, Kenealy, Beyer, Zhai, Tsitsulin, Busa-Fekete, Feng, Sachdeva, Coleman, Gao, Mustafa, Barr, Parisotto, Tian, Eyal, Cherry, Peter, Sinopalnikov, Bhupatiraju, Agarwal, Kazemi, Malkin, Kumar, Vilar, Brusilovsky, Luo, Steiner, Friesen, Sharma, Sharma, Gilady, Goedeckemeyer, Saade, Feng, Kolesnikov, Bendebury, Abdagic, Vadi, György, Pinto, Das, Bapna, Miech, Yang, Paterson, Shenoy, Chakrabarti, Piot, Wu, Shahriari, Petrini, Chen, Lan, Choquette-Choo, Carey, Brick, Deutsch, Eisenbud, Cattle, Cheng, Paparas, Sreepathihalli, Reid, Tran, Zelle, Noland, Huizenga, Kharitonov, Liu, Amirkhanyan, Cameron, Hashemi, Klimczak-Plucińska, Singh, Mehta, Lehri, Hazimeh, Ballantyne, Szpektor, Nardini, Pouget-Abadie, Chan, Stanton, Wieting, Lai, Orbay, Fernandez, Newlan, yeong Ji, Singh, Black, Yu, Hui, Vodrahalli, Greff, Qiu,
  Valentine, Coelho, Ritter, Hoffman, Watson, Chaturvedi, Moynihan, Ma, Babar, Noy, Byrd, Roy, Momchev, Chauhan, Sachdeva, Bunyan, Botarda, Caron, Rubenstein, Culliton, Schmid, Sessa, Xu, Stanczyk, Tafti, Shivanna, Wu, Pan, Rokni, Willoughby, Vallu, Mullins, Jerome, Smoot, Girgin, Iqbal, Reddy, Sheth, Põder, Bhatnagar, Panyam, Eiger, Zhang, Liu, Yacovone, Liechty, Kalra, Evci, Misra, Roseberry, Feinberg, Kolesnikov, Han, Kwon, Chen, Chow, Zhu, Wei, Egyed, Cotruta, Giang, Kirk, Rao, Black, Babar, Lo, Moreira, Martins, Sanseviero, Gonzalez, Gleicher, Warkentin, Mirrokni, Senter, Collins, Barral, Ghahramani, Hadsell, Matias, Sculley, Petrov, Fiedel, Shazeer, Vinyals, Dean, Hassabis, Kavukcuoglu, Farabet, Buchatskaya, Alayrac, Anil, Dmitry, Lepikhin, Borgeaud, Bachem, Joulin, Andreev, Hardin, Dadashi, and Hussenot}]{gemmateam2025gemma3technicalreport}
Gemma Team, Aishwarya Kamath, Johan Ferret, Shreya Pathak, Nino Vieillard, Ramona Merhej, Sarah Perrin, Tatiana Matejovicova, Alexandre Ramé, Morgane Rivière, Louis Rouillard, Thomas Mesnard, Geoffrey Cideron, Jean bastien Grill, Sabela Ramos, Edouard Yvinec, Michelle Casbon, Etienne Pot, Ivo Penchev, Gaël Liu, Francesco Visin, Kathleen Kenealy, Lucas Beyer, Xiaohai Zhai, Anton Tsitsulin, Robert Busa-Fekete, Alex Feng, Noveen Sachdeva, Benjamin Coleman, Yi~Gao, Basil Mustafa, Iain Barr, Emilio Parisotto, David Tian, Matan Eyal, Colin Cherry, Jan-Thorsten Peter, Danila Sinopalnikov, Surya Bhupatiraju, Rishabh Agarwal, Mehran Kazemi, Dan Malkin, Ravin Kumar, David Vilar, Idan Brusilovsky, Jiaming Luo, Andreas Steiner, Abe Friesen, Abhanshu Sharma, Abheesht Sharma, Adi~Mayrav Gilady, Adrian Goedeckemeyer, Alaa Saade, Alex Feng, Alexander Kolesnikov, Alexei Bendebury, Alvin Abdagic, Amit Vadi, András György, André~Susano Pinto, Anil Das, Ankur Bapna, Antoine Miech, Antoine Yang, Antonia Paterson, Ashish
  Shenoy, Ayan Chakrabarti, Bilal Piot, Bo~Wu, Bobak Shahriari, Bryce Petrini, Charlie Chen, Charline~Le Lan, Christopher~A. Choquette-Choo, CJ~Carey, Cormac Brick, Daniel Deutsch, Danielle Eisenbud, Dee Cattle, Derek Cheng, Dimitris Paparas, Divyashree~Shivakumar Sreepathihalli, Doug Reid, Dustin Tran, Dustin Zelle, Eric Noland, Erwin Huizenga, Eugene Kharitonov, Frederick Liu, Gagik Amirkhanyan, Glenn Cameron, Hadi Hashemi, Hanna Klimczak-Plucińska, Harman Singh, Harsh Mehta, Harshal~Tushar Lehri, Hussein Hazimeh, Ian Ballantyne, Idan Szpektor, Ivan Nardini, Jean Pouget-Abadie, Jetha Chan, Joe Stanton, John Wieting, Jonathan Lai, Jordi Orbay, Joseph Fernandez, Josh Newlan, Ju~yeong Ji, Jyotinder Singh, Kat Black, Kathy Yu, Kevin Hui, Kiran Vodrahalli, Klaus Greff, Linhai Qiu, Marcella Valentine, Marina Coelho, Marvin Ritter, Matt Hoffman, Matthew Watson, Mayank Chaturvedi, Michael Moynihan, Min Ma, Nabila Babar, Natasha Noy, Nathan Byrd, Nick Roy, Nikola Momchev, Nilay Chauhan, Noveen Sachdeva, Oskar
  Bunyan, Pankil Botarda, Paul Caron, Paul~Kishan Rubenstein, Phil Culliton, Philipp Schmid, Pier~Giuseppe Sessa, Pingmei Xu, Piotr Stanczyk, Pouya Tafti, Rakesh Shivanna, Renjie Wu, Renke Pan, Reza Rokni, Rob Willoughby, Rohith Vallu, Ryan Mullins, Sammy Jerome, Sara Smoot, Sertan Girgin, Shariq Iqbal, Shashir Reddy, Shruti Sheth, Siim Põder, Sijal Bhatnagar, Sindhu~Raghuram Panyam, Sivan Eiger, Susan Zhang, Tianqi Liu, Trevor Yacovone, Tyler Liechty, Uday Kalra, Utku Evci, Vedant Misra, Vincent Roseberry, Vlad Feinberg, Vlad Kolesnikov, Woohyun Han, Woosuk Kwon, Xi~Chen, Yinlam Chow, Yuvein Zhu, Zichuan Wei, Zoltan Egyed, Victor Cotruta, Minh Giang, Phoebe Kirk, Anand Rao, Kat Black, Nabila Babar, Jessica Lo, Erica Moreira, Luiz~Gustavo Martins, Omar Sanseviero, Lucas Gonzalez, Zach Gleicher, Tris Warkentin, Vahab Mirrokni, Evan Senter, Eli Collins, Joelle Barral, Zoubin Ghahramani, Raia Hadsell, Yossi Matias, D.~Sculley, Slav Petrov, Noah Fiedel, Noam Shazeer, Oriol Vinyals, Jeff Dean, Demis Hassabis,
  Koray Kavukcuoglu, Clement Farabet, Elena Buchatskaya, Jean-Baptiste Alayrac, Rohan Anil, Dmitry, Lepikhin, Sebastian Borgeaud, Olivier Bachem, Armand Joulin, Alek Andreev, Cassidy Hardin, Robert Dadashi, and Léonard Hussenot. 2025.
\newblock \href {http://arxiv.org/abs/2503.19786} {Gemma 3 technical report}.

\bibitem[{Team(2024)}]{Codestral}
Mistral~AI Team. 2024.
\newblock \href {https://mistral.ai/news/codestral/} {Codestral: Hello, world!}

\bibitem[{Team(2025)}]{qwq32b}
Qwen Team. 2025.
\newblock \href {https://qwenlm.github.io/blog/qwq-32b/} {Qwq-32b: Embracing the power of reinforcement learning}.

\bibitem[{Wang et~al.(2024)Wang, Yang, Wang, Huang, Chu, Song, Zhang, Chen, and Ma}]{Wang2024TESTEVALBL}
Wenhan Wang, Chenyuan Yang, Zhijie Wang, Yuheng Huang, Zhaoyang Chu, Da~Song, Lingming Zhang, An~Ran Chen, and Lei Ma. 2024.
\newblock \href {https://api.semanticscholar.org/CorpusID:270357873} {Testeval: Benchmarking large language models for test case generation}.
\newblock \emph{ArXiv}, abs/2406.04531.

\bibitem[{Yang et~al.(2025)Yang, Li, Yang, Zhang, Hui, Zheng, Yu, Gao, Huang, Lv, Zheng, Liu, Zhou, Huang, Hu, Ge, Wei, Lin, Tang, Yang, Tu, Zhang, Yang, Yang, Zhou, Zhou, Lin, Dang, Bao, Yang, Yu, Deng, Li, Xue, Li, Zhang, Wang, Zhu, Men, Gao, Liu, Luo, Li, Tang, Yin, Ren, Wang, Zhang, Ren, Fan, Su, Zhang, Zhang, Wan, Liu, Wang, Cui, Zhang, Zhou, and Qiu}]{qwen3}
An~Yang, Anfeng Li, Baosong Yang, Beichen Zhang, Binyuan Hui, Bo~Zheng, Bowen Yu, Chang Gao, Chengen Huang, Chenxu Lv, Chujie Zheng, Dayiheng Liu, Fan Zhou, Fei Huang, Feng Hu, Hao Ge, Haoran Wei, Huan Lin, Jialong Tang, Jian Yang, Jianhong Tu, Jianwei Zhang, Jianxin Yang, Jiaxi Yang, Jing Zhou, Jingren Zhou, Junyang Lin, Kai Dang, Keqin Bao, Kexin Yang, Le~Yu, Lianghao Deng, Mei Li, Mingfeng Xue, Mingze Li, Pei Zhang, Peng Wang, Qin Zhu, Rui Men, Ruize Gao, Shixuan Liu, Shuang Luo, Tianhao Li, Tianyi Tang, Wenbiao Yin, Xingzhang Ren, Xinyu Wang, Xinyu Zhang, Xuancheng Ren, Yang Fan, Yang Su, Yichang Zhang, Yinger Zhang, Yu~Wan, Yuqiong Liu, Zekun Wang, Zeyu Cui, Zhenru Zhang, Zhipeng Zhou, and Zihan Qiu. 2025.
\newblock Qwen3 technical report.
\newblock \emph{arXiv preprint arXiv:2505.09388}.

\bibitem[{Yang et~al.(2024{\natexlab{a}})Yang, Yang, Hui, Zheng, Yu, Zhou, Li, Li, Liu, Huang, Dong, Wei, Lin, Tang, Wang, Yang, Tu, Zhang, Ma, Xu, Zhou, Bai, He, Lin, Dang, Lu, Chen, Yang, Li, Xue, Ni, Zhang, Wang, Peng, Men, Gao, Lin, Wang, Bai, Tan, Zhu, Li, Liu, Ge, Deng, Zhou, Ren, Zhang, Wei, Ren, Fan, Yao, Zhang, Wan, Chu, Cui, Zhang, and Fan}]{Yang2024Qwen2TR}
An~Yang, Baosong Yang, Binyuan Hui, Bo~Zheng, Bowen Yu, Chang Zhou, Chengpeng Li, Chengyuan Li, Dayiheng Liu, Fei Huang, Guanting Dong, Haoran Wei, Huan Lin, Jialong Tang, Jialin Wang, Jian Yang, Jianhong Tu, Jianwei Zhang, Jianxin Ma, Jin Xu, Jingren Zhou, Jinze Bai, Jinzheng He, Junyang Lin, Kai Dang, Keming Lu, Ke-Yang Chen, Kexin Yang, Mei Li, Min Xue, Na~Ni, Pei Zhang, Peng Wang, Ru~Peng, Rui Men, Ruize Gao, Runji Lin, Shijie Wang, Shuai Bai, Sinan Tan, Tianhang Zhu, Tianhao Li, Tianyu Liu, Wenbin Ge, Xiaodong Deng, Xiaohuan Zhou, Xingzhang Ren, Xinyu Zhang, Xipin Wei, Xuancheng Ren, Yang Fan, Yang Yao, Yichang Zhang, Yunyang Wan, Yunfei Chu, Zeyu Cui, Zhenru Zhang, and Zhi-Wei Fan. 2024{\natexlab{a}}.
\newblock \href {https://api.semanticscholar.org/CorpusID:271212307} {Qwen2 technical report}.
\newblock \emph{ArXiv}, abs/2407.10671.

\bibitem[{Yang et~al.(2024{\natexlab{b}})Yang, Yang, Zhang, Hui, Zheng, Yu, Li, Liu, Huang, Dong, Wei, Lin, Yang, Tu, Zhang, Yang, Yang, Zhou, Lin, Dang, Lu, Bao, Yang, Yu, Li, Xue, Zhang, Zhu, Men, Lin, Li, Xia, Ren, Ren, Fan, Su, Zhang, Wan, Liu, Cui, Zhang, Qiu, Quan, and Wang}]{Yang2024Qwen25TR}
Qwen~An Yang, Baosong Yang, Beichen Zhang, Binyuan Hui, Bo~Zheng, Bowen Yu, Chengyuan Li, Dayiheng Liu, Fei Huang, Guanting Dong, Haoran Wei, Huan Lin, Jian Yang, Jianhong Tu, Jianwei Zhang, Jianxin Yang, Jiaxin Yang, Jingren Zhou, Junyang Lin, Kai Dang, Keming Lu, Keqin Bao, Kexin Yang, Le~Yu, Mei Li, Mingfeng Xue, Pei Zhang, Qin Zhu, Rui Men, Runji Lin, Tianhao Li, Tingyu Xia, Xingzhang Ren, Xuancheng Ren, Yang Fan, Yang Su, Yi-Chao Zhang, Yunyang Wan, Yuqi Liu, Zeyu Cui, Zhenru Zhang, Zihan Qiu, Shanghaoran Quan, and Zekun Wang. 2024{\natexlab{b}}.
\newblock \href {https://api.semanticscholar.org/CorpusID:274859421} {Qwen2.5 technical report}.
\newblock \emph{ArXiv}, abs/2412.15115.

\bibitem[{Yu et~al.(2024{\natexlab{a}})Yu, Zhao, Cohan, and Zhang}]{yu2024humanevalprombpppro}
Zhaojian Yu, Yilun Zhao, Arman Cohan, and Xiao-Ping Zhang. 2024{\natexlab{a}}.
\newblock \href {http://arxiv.org/abs/2412.21199} {Humaneval pro and mbpp pro: Evaluating large language models on self-invoking code generation}.

\bibitem[{Yu et~al.(2024{\natexlab{b}})Yu, Zhao, Cohan, and Zhang}]{Yu2024HumanEvalPA}
Zhaojian Yu, Yilun Zhao, Arman Cohan, and Xiao-Ping Zhang. 2024{\natexlab{b}}.
\newblock \href {https://api.semanticscholar.org/CorpusID:275134345} {Humaneval pro and mbpp pro: Evaluating large language models on self-invoking code generation}.
\newblock \emph{ArXiv}, abs/2412.21199.

\bibitem[{Zeng et~al.(2025)Zeng, Jiang, Wang, Nie, Chen, and Chen}]{Zeng2025ACECODERAC}
Huaye Zeng, Dongfu Jiang, Haozhe Wang, Ping Nie, Xiaotong Chen, and Wenhu Chen. 2025.
\newblock \href {https://api.semanticscholar.org/CorpusID:276107488} {Acecoder: Acing coder rl via automated test-case synthesis}.

\bibitem[{Zhang et~al.(2023)Zhang, Wang, Xia, Wang, and Li}]{Zhang2023ALGOSA}
Kexun Zhang, Danqing Wang, Jingtao Xia, William~Yang Wang, and Lei Li. 2023.
\newblock \href {https://api.semanticscholar.org/CorpusID:258865731} {Algo: Synthesizing algorithmic programs with generated oracle verifiers}.
\newblock \emph{ArXiv}, abs/2305.14591.

\end{thebibliography}

\appendix

\clearpage
\section{Related Work}
\subsection{Line and Branch Coverage}
\label{app:related-work}

In the realm of software testing, two pervasive metrics are Line Coverage and Branch Coverage. These are often used to evaluate the adequacy of test cases in executing program code.

Specifically, \textbf{Line Coverage} measures the percentage of lines of code that have been executed by a set of test cases. It provides insight into which lines of the codebase are actually executed during testing, aiming to ensure that all parts of the code are tested at least once.
\textbf{Branch Coverage} takes a more granular approach by focusing on the control structures within the code, such as if statements and switch cases. It evaluates whether each possible branch (i.e., each path through a control structure) has been executed. This metric ensures that all possible execution paths are tested.

While these metrics are invaluable in traditional software testing, they fall short in the context of competitive algorithm problems for several reasons:
(1) Algorithm Complexity and Diversity: Competitive algorithm problems often involve complex data structures and intricate algorithmic logic that cannot be fully represented by simple line or branch execution. The focus is on the correctness and efficiency of the algorithm, rather than merely executing each line or branch of code.
(2) Outcome-Oriented Nature: The primary goal in algorithm competitions is to solve problems correctly and efficiently, not just to achieve high code coverage. An algorithm may achieve high line and branch coverage but still fail to solve the problem correctly or efficiently.
(3) Diversity of Test Cases: Algorithm competition problems require testing against a wide variety of edge cases and specific inputs. The generation and evaluation of these test cases extend beyond the scope of simple line and branch coverage metrics, which may not adequately reflect the comprehensiveness of the test cases in ensuring algorithmic correctness and robustness.
Traditional Line and Branch Coverage metrics may be inadequate in the context of algorithmic problems. Therefore, we propose two new tasks, Fault Coverage and Fault Exposure, along with corresponding evaluation metrics.

\section{\ours Benchmark}
We provide a detailed explanation of the problems and human-written solutions within the dataset.

\begin{figure}[!t]
\centering
\includegraphics[width=0.7\linewidth]{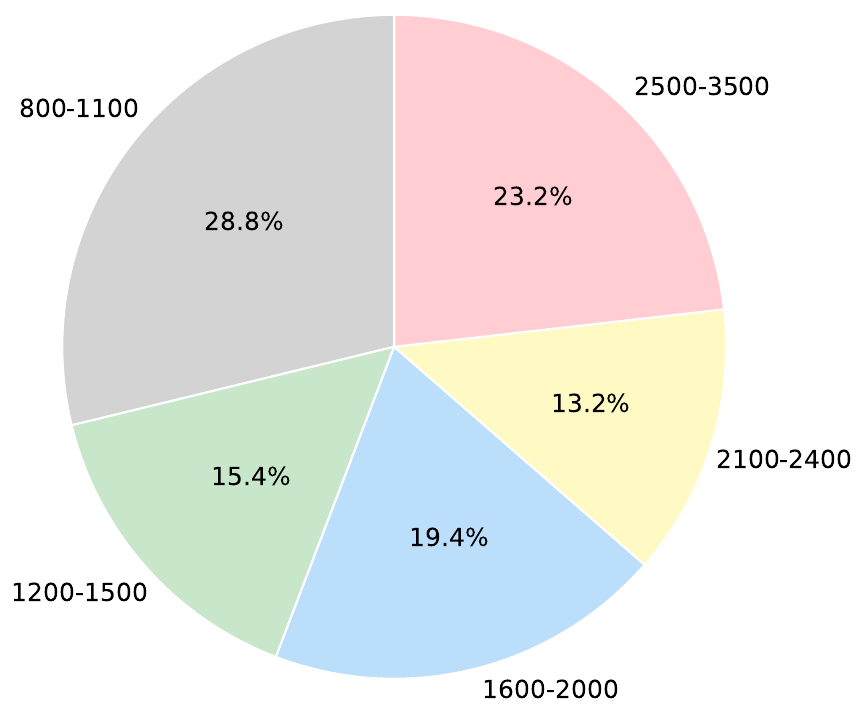}
\caption{
Distribution of Problem Difficulty Levels.
}
\label{fig:problem-difficulty-distribution}
\end{figure}

\subsection{Problems}

Each problem sourced from the \cf platform comprises several key elements: 1) title, 2) time limit, 3) memory limit, 4) problem description, 5) input format, 6) output format, 7) test case examples, and 8) optional note. We utilize all of this data to form the \texttt{problem\_description} string, which acts as the input for the LLM.

Additionally, we analyze the distribution of problem difficulty ratings, which is illustrated in Figure \autoref{fig:problem-difficulty-distribution}.

\subsection{Human-written Solutions}
\label{app:human-written-solutions}

To ensure both the representativeness and diversity of error patterns in our benchmark, we developed a comprehensive dataset collection and sampling pipeline.

For each selected Codeforces problem, we first crawled the complete submission logs to collect a representative set of user-submitted incorrect solutions. Specifically, for each problem, we initially sampled 100 incorrect solutions for each of the three major programming languages (C++, Python, and Java), resulting in a preliminary pool. We then applied multiple rounds of filtering and cleaning to ensure quality and diversity. This process yielded a final set of 118,611 human-written incorrect solutions across 500 algorithmic problems, amounting to an average of 237 solutions per problem. These submissions reflect genuine programming errors from a diverse pool of users, capturing a broad spectrum of error types and difficulty levels observed in real-world programming scenarios.

We imposed strict criteria on the sampled solutions: each must be semantically valid and executable, passing compilation and basic test cases without syntax or runtime errors, and failing only under specific, non-trivial input conditions. This design ensures our benchmark targets input-sensitive faults—precisely those that require sophisticated and diverse test input generation to detect.

To construct a manageable yet representative subset for Task 2, we performed stratified sampling for each problem. We began by analyzing the distribution of incorrect solutions by error type and difficulty, which we defined based on the index of the first failed test case. Guided by this analysis, we sampled 20 incorrect solutions per problem, ensuring balanced representation across three major programming languages and maintaining proportional coverage of both error types and difficulty levels.

Codeforces problems typically include between a dozen and over two hundred test cases, each comprising a set of inputs and expected outputs. For a given submission, the verdict ``Wrong answer on test 5'' indicates that the solution passed the first four test cases but failed on the fifth. The index of the first failed test case thus serves as a crucial signal for assessing both the difficulty of a solution and the effectiveness of generated test cases.

Based on this index, we categorize solutions into three levels of difficulty: \emph{Easy}, \emph{Medium}, and \emph{Hard}. Specifically, we sort all human-written solutions for each problem by the index of the first error, assigning the bottom 40\% to \emph{Easy}, the middle 30\% to \emph{Medium}, and the top 30\% to \emph{Hard}.

\begin{figure*}[!t]
\centering
\includegraphics[width=0.7\linewidth]{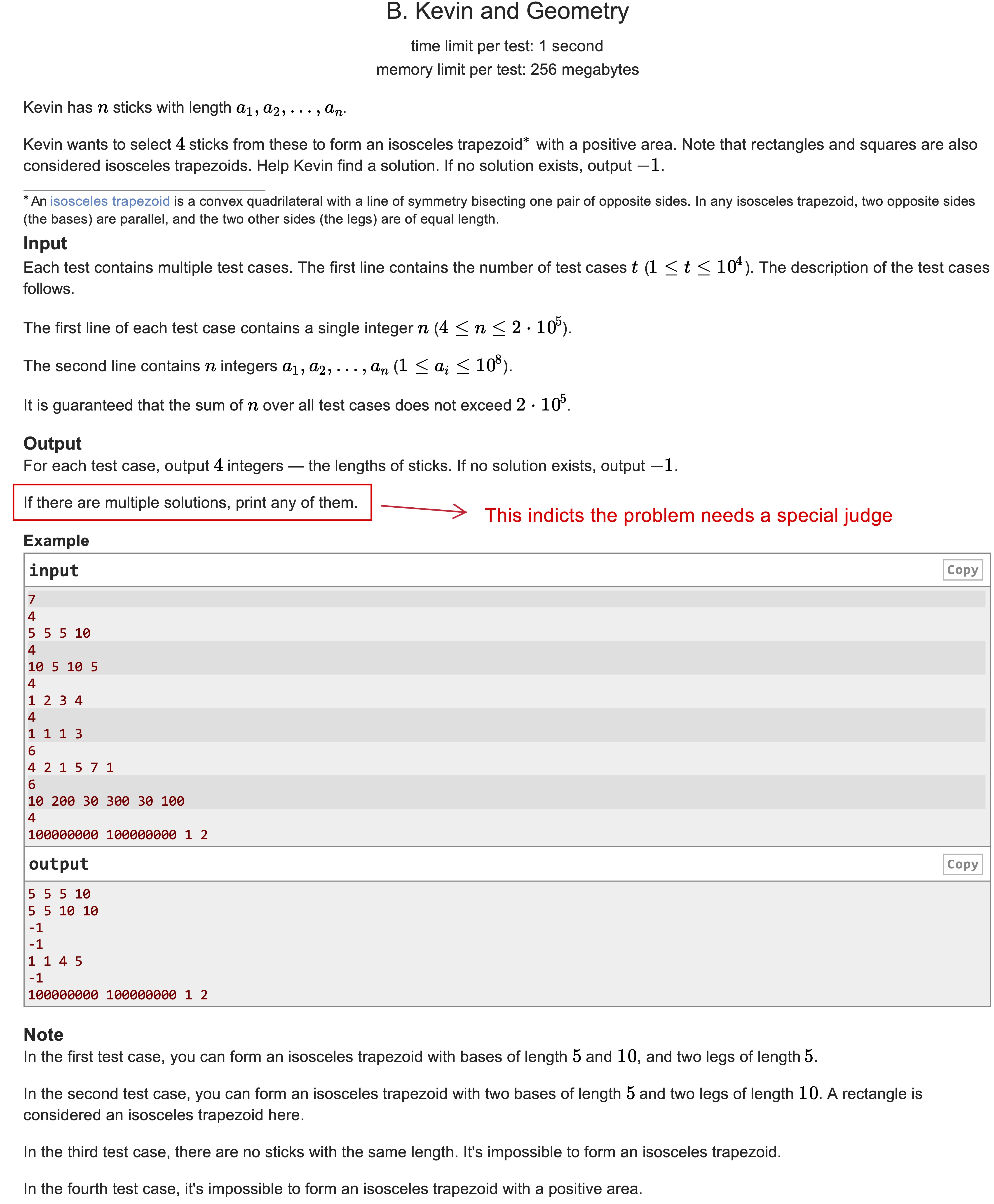}
\caption{
An example of a problem that needs a special judge.
}
\label{fig:problem-special-judge}
\end{figure*}

\subsection{Special Judge} \label{app:judge}
In competitive programming platforms like \cf, certain problems permit multiple valid test outputs for a single test input. To validate such outputs, a special judge is employed. This custom code evaluates the correctness of each output, as straightforward comparison to a reference output is inadequate due to the problem's complexity. \autoref{fig:problem-special-judge} illustrates a problem that necessitates a special judge.

For accurate offline evaluation, we excluded all problems requiring a special judge. Such problems can cause inconsistent assessments since they allow multiple correct outputs for the same input.

\subsection{Multiple Test Cases}
In competitive programming platforms (especially Codeforces), multiple test cases within a single test input are a standard feature. As illustrated in \autoref{fig:problem-special-judge}, problem input specifications often begin with instructions such as ``Each test contains multiple test cases'' emphasizing the expectation that solutions correctly process a batch of cases in one execution. Through manual inspection of our dataset, we verified that 439 out of the 500 problems inherently require handling multiple test cases. To ensure comprehensive test coverage, we designed our prompt in Appendix \ref{app:prompt} to explicitly guide LLMs in generating diverse test cases in a single test input.

\onecolumn
\section{Experiment Setup}
\subsection{Evaluated Model Configuration}\label{app:model_info}
\autoref{tab:model_list} details the configuration of each evaluated model. Across all experiments, the temperature is set to 1.0 to ensure diversity in the LLM-generated test cases. The maximum output length is generally configured to 2048 tokens, which suffices for most standard models. However, for reasoning models like QwQ-32B and R1-Distill-Qwen-32B, this maximum output length is extended to 18,000 tokens to accommodate their long CoT reasoning mechanisms. All inference processes are conducted on two NVIDIA A100-80G GPUs.
\begin{table*}[h]
    \centering
    \small
    \begin{tabular}{l|l|l}
        \toprule
        \textbf{Model} & \textbf{Citation} & \textbf{Version} \\ \midrule

        GPT-4.1 & \citet{openai2025gpt41} & gpt-4.1-2025-04-14 \\
        GPT-4.1-mini & \citet{openai2025gpt41} & gpt-4.1-mini-2025-04-14\\
        GPT-4o & \citet{openai2024gpt4o} & gpt-4o-2024-11-20 \\
        \midrule
    
        Qwen3-8B & \citet{qwen3} & \href{https://huggingface.co/Qwen/Qwen3-8B}{Qwen/Qwen3-8B} \\
        Qwen3-32B & \citet{qwen3} & \href{https://huggingface.co/Qwen/Qwen3-32B}{Qwen/Qwen3-32B} \\
        R1-Distill-Qwen-32B & \citet{DeepSeekAI2025DeepSeekR1IR} & \href{https://huggingface.co/deepseek-ai/DeepSeek-R1-Distill-Qwen-32B}{deepseek-ai/DeepSeek-R1-Distill-Qwen-32B} \\
        QwQ-32B & \citet{qwq32b} & \href{https://huggingface.co/Qwen/QwQ-32B}{Qwen/QwQ-32B} \\

        Qwen2.5-7B & \citet{Yang2024Qwen2TR} & \href{Qwen/Qwen2.5-7B-Instruct}{Qwen/Qwen2.5-7B-Instruct} \\
        Qwen2.5-Coder-7B & \citet{Hui2024Qwen25CoderTR} & \href{https://huggingface.co/Qwen/Qwen2.5-Coder-7B-Instruct}{Qwen/Qwen2.5-Coder-7B-Instruct} \\
        Qwen2.5-32B & \citet{Yang2024Qwen25TR} & \href{https://huggingface.co/Qwen/Qwen2.5-32B-Instruct}{Qwen/Qwen2.5-32B-Instruct} \\
        Qwen2.5-Coder-32B & \citet{Hui2024Qwen25CoderTR} & \href{https://huggingface.co/Qwen/Qwen2.5-Coder-32B-Instruct}{Qwen/Qwen2.5-Coder-32B-Instruct} \\
        Qwen2.5-72B & \citet{Yang2024Qwen25TR} & \href{https://huggingface.co/Qwen/Qwen2.5-72B-Instruct}{Qwen/Qwen2.5-72B-Instruct} \\
        
        Llama-3.1-70B & \citet{Dubey2024TheL3} & \href{https://huggingface.co/meta-llama/Llama-3.1-70B-Instruct}{meta-llama/Llama-3.1-70B-Instruct} \\
        Llama-3.3-70B & \citet{Dubey2024TheL3} & \href{https://huggingface.co/meta-llama/Llama-3.3-70B-Instruct}{meta-llama/Llama-3.3-70B-Instruct} \\

        Mistral-Small-24B & \citet{Jiang2023Mistral7} & \href{https://huggingface.co/mistralai/Mistral-Small-24B-Instruct-2501}{mistralai/Mistral-Small-24B-Instruct-2501} \\
        Codestral-22B & \citet{Codestral} & \href{https://huggingface.co/mistralai/Codestral-22B-v0.1}{mistralai/Codestral-22B-v0.1} \\

        Gemma-3-12B & \citet{gemmateam2025gemma3technicalreport} & \href{https://huggingface.co/google/gemma-3-12b-it}{google/gemma-3-12b-it} \\
        Gemma-3-27B & \citet{gemmateam2025gemma3technicalreport} & \href{https://huggingface.co/google/gemma-3-27b-it}{google/gemma-3-27b-it} \\
        Seed-Coder-8B & \citet{seedcoder} & \href{https://huggingface.co/ByteDance-Seed/Seed-Coder-8B-Instruct}{ByteDance-Seed/Seed-Coder-8B-Instruct}\\
\bottomrule

    \end{tabular}
    \caption{Model List.}
    \label{tab:model_list}
\end{table*}

\subsection{CoT and Direct Output Prompts}\label{app:prompt}

\begin{figure}[h]
    \begin{tcolorbox}[colback=black!3!white, colframe=black!70!white, title=The Chain-of-Thought Prompt in Task1, fontupper=\footnotesize, fonttitle=\footnotesize]
        \textbf{Task:} \\
        Generate a challenging test input for the algorithm problem: \\
        \textcolor{blue}{\{problem\_description\}} \\

        \textbf{Instructions:} \\
        - Focus on edge cases or scenarios that maximize the failure probability in faulty solutions. \\
        - Due to the output length limit, you should generate a small-scale test input that is complete and valid. \\
        - Output the test input directly, not code to generate it. \\

        \textbf{Output format:} 
        \begin{verbatim}
```plaintext
{test input}
```
        \end{verbatim}
        
        \textbf{Think step by step.}
    \end{tcolorbox}
    \caption{The Chain-of-Thought prompt used in Task1.}
\end{figure}

\begin{figure}[h]
    \begin{tcolorbox}[colback=black!3!white, colframe=black!70!white, title=The Direct Output prompt in Task1, fontupper=\footnotesize, fonttitle=\footnotesize]
        \textbf{Task:} \\
        Generate a challenging test input for the algorithm problem: \\
        \textcolor{blue}{\{problem\_description\}} \\

        \textbf{Instructions:} \\
        - Focus on edge cases or scenarios that maximize the failure probability in faulty solutions. \\
        - Due to the output length limit, you should generate a small-scale test input that is complete and valid. \\
        - Output the test input directly, not code to generate it. \\

        \textbf{Output format:} 
        \begin{verbatim}
```plaintext
{test input}
```
        \end{verbatim}
        
        \textbf{Only output the test input, no explanations.}
    \end{tcolorbox}
    \caption{The Direct Output prompt used in Task1.}
\end{figure}

\begin{figure}[h]
    \begin{tcolorbox}[colback=black!3!white, colframe=black!70!white, title=The Chain-of-Thought prompt in Task2, fontupper=\footnotesize, fonttitle=\footnotesize]
        \textbf{Task:} \\
        Generate a challenging test input that exposes the bug in the buggy code of the algorithm problem: \\
        Algorithm Problem: \textcolor{blue}{\{problem\_description\}} \\
        Buggy Code: \textcolor{blue}{\{buggy\_code\}} \\

        \textbf{Instructions:} \\
        - Focus on edge cases or scenarios that maximize the failure probability in faulty solutions. \\
        - Due to the output length limit, you should generate a small-scale test input that is complete and valid. \\
        - Output the test input directly, not code to generate it. \\

        \textbf{Output format:} 
        \begin{verbatim}
```plaintext
{test input}
```
        \end{verbatim}
      
        \textbf{Think step by step.}
    \end{tcolorbox}
    \caption{The Chain-of-Thought prompt used in Task2.}
\end{figure}

\begin{figure}[h]
    \begin{tcolorbox}[colback=black!3!white, colframe=black!70!white, title=The Direct Output prompt in Task2, fontupper=\footnotesize, fonttitle=\footnotesize]
        \textbf{Task:} \\
        Generate a challenging test input that exposes the bug in the buggy code of the algorithm problem: \\
        Algorithm Problem: \textcolor{blue}{\{problem\_description\}} \\
        Buggy Code: \textcolor{blue}{\{buggy\_code\}} \\

        \textbf{Instructions:} \\
        - Focus on edge cases or scenarios that maximize the failure probability in faulty solutions. \\
        - Due to the output length limit, you should generate a small-scale test input that is complete and valid. \\
        - Output the test input directly, not code to generate it. \\

        \textbf{Output format:} 
        \begin{verbatim}
```plaintext
{test input}
```
        \end{verbatim}
      
        \textbf{Only output the test input, no explanations.}
    \end{tcolorbox}
    \caption{The Direct Output prompt used in Task2.}
\end{figure}

\end{document}